\documentclass[]{spie}  %>>> use for US letter paper
\usepackage[]{graphicx}
\title{ POLARIX: a small mission of x-ray polarimetry}

\author{ Enrico Costa\supit{a},  Ronaldo Bellazzini\supit{b},
Paolo  Soffitta\supit{a}, Fabio  Muleri\supit{a},  Marco
Feroci\supit{a}, Massimo Frutti\supit{a}, Marcello
Mastropietro\supit{a}, Luigi Pacciani\supit{a}, Alda
Rubini\supit{a}, Ennio Morelli\supit{c}, Luca Baldini\supit{b},
Francesco Bitti\supit{b}, Alessandro Brez\supit{b}, Francesco
Cavalca\supit{b}, Luca Latronico\supit{b},  Marco Maria
Massai\supit{b},  Nicola Omodei\supit{b}, Michele
Pinchera\supit{b}, Carmelo Sgr{\'o}\supit{b}, Gloria
Spandre\supit{b}, Giorgio Matt\supit{d}, Giuseppe Cesare
Perola\supit{d}, Guido Chincarini\supit{e}\supit{f}, Oberto
Citterio\supit{e}, Gianpiero Tagliaferri\supit{e}, Giovanni
Pareschi\supit{e}, Vincenzo Cotroneo\supit{e} \skiplinehalf
\supit{a} Istituto di Astrofisica Spaziale e Fisica Cosmica, Via
del Fosso del Cavaliere 100, I-00133 Roma, Italy;
\\
\supit{b} Istituto Nazionale di Fisica Nucleare, Largo B.
Pontecorvo 3, I-56127 Pisa,  Italy
\\
\supit{c} Istituto di
Astrofisica Spaziale e Fisica Cosmica, Via Gobetti 101, I-40129
Bologna, Italy
\\
\supit{d} Universita' degli Studi di Roma 3, Via della Vasca
Navale 84, I-00146 Roma, Italy
\\
\supit{e} Osservatorio Astronomico di Brera, Via E. Bianchi 46,
I-23807 Merate (LC), Italy
\\\supit{f} Universita' Milano Bicocca, Piazza delle Scienze 3, I-20126 Milano,
Italy }

\authorinfo{Send correspondence to Enrico Costa:
E-mail:enrico.costa@iasf-roma.inaf.it, Telephone: +39 0649934004}
\begin{document}
  \maketitle

%%%%%%%%%%%%%%%%%%%%%%%%%%%%%%%%%%%%%%%%%%%%%%%%%%%%%%%%%%%%%
\begin{abstract}
X-Ray Polarimetry can be now performed by using a Micro Pattern
Gas Chamber in the focus of a telescope. It requires large area
optics for most important scientific targets. But since the
technique is additive a dedicated mission with a cluster of small
telescopes can perform many important measurements and bridge the
40 year gap between OSO-8 data and future big telescopes such as
XEUS. POLARIX has been conceived as such a pathfinder. It is a
Small Satellite based on the optics of JET-X. Two telescopes are
available in flight configuration and three more can be easily
produced starting from the available superpolished mandrels. We
show the capabilities of such a cluster of telescopes each
equipped with a focal plane photoelectric polarimeter and discuss
a few alternative solutions.
\end{abstract}

%>>>> Include a list of keywords after the abstract


%%%%%%%%%%%%%%%%%%%%%%%%%%%%%%%%%%%%%%%%%%%%%%%%%%%%%%%%%%%%%
\section{Polarimetry: Big Hopes Meagre results.}
\label{sect:intro}  % \label{} allows reference to this section
A long term theoretical analysis has foreseen the possibility to
test models of X-ray sources and to derive relevant parameters of
the models themselves by measuring the linear polarization of
X-rays. In the first years the polarization deriving from the
emission process was outlined. On 1985 Sunyaev and  and Titarchuk
demonstrated that in a hot accretion disk the scattered photons
have an energy much higher than the original photons and thence
the Chandrasekhar limit of 11.7 $\%$ could be exceeded. In 1988
Meszaros et al. stated "Polarimetry would add to energy and time
two further observable quantities (the amount and the angle of
polarization) constraining any model and interpretation: a
theoretical/observational break-through".

This very attractive perspective pushed some investigators to
develop instrumentation aimed to perform this measurements. This
was mainly based on Bragg Diffraction at 45$^o$ and on
Compton/Thomson scattering in a certain range of angles around
90$^o$. Bragg crystal is an excellent analyzer of polarization
(100$\%$ modulation) but is effective in a very narrow interval
around the energy that fulfils the Bragg condition for that
particular crystal and that particular angle. A flat Bragg crystal
at 45$^o$ set before the focal plane of an X-ray telescope acts as
a newtonian secondary mirror and a detector in the secondary focal
plane detects the source in the thin band defined by the Bragg
(plus higher orders if in the band). The whole is rotated and the
source brightness is modulated depending on the amount of
polarization.

The scattering polarimetry is effective only when scattering
exceeds photoabsorption. A well of detectors surrounds the
scatterer and the polarization is measured by the angular
distribution of the out-coming photons. This technique can be used
by means of large experiments with a distributed multiplicity of
detectors and scatterers and many experiments based on this
principle have been proposed and some are on the
way\cite{Legere05}. Scattering polarimetry can also be performed
in the focus of a telescope with some substantial limitations:
even with lithium the method is only effective above 5 keV and
this is mismatched with the bandpass of most telescopes; the
angular distribution strongly depends on the impact point and the
measurement is severely affected by systematics; all the detector
set (and not only a few pixels) is involved in the measurement so
that the improvement of signal to noise ratio introduced by the
use of the telescope is essentially lost.

The pioneering result achieved by the Columbia University team was
the measurement of polarization of Crab Nebula first with a
rocket\cite{Novick72} then with a Bragg polarimeter on-board OSO-8
satellite\cite{Weisskopf76}. This result was of extreme relevance
but the overall throughput of X-ray polarimetry based on
conventional techniques was considered too meagre to support the
inclusion of a polarimeter onboard further X-ray missions.
Moreover theoretical analysis suggested that a polarimeter capable
to perform polarimetry at few $\%$ level also on relatively faint
sources was needed for most of the scientific objectives. The only
mission with a focal plane polarimeter (both Bragg and Compton)
was the Spectrum X-Gamma\cite{kaaret89} that sinked in the general
collapse of the soviet system.

A third physical process that can analyze the polarization is
photoelectric absorption: s-photoelectrons are ejected according
to a cos$^{2}$ distribution, with respect to the electric field.
This is a potentially ideal analyzer of polarization and many
teams tried, also in early times, to build a device based on this
effect. The difficulty is that electrons penetrate much less than
photons, and a finely subdivided detector is needed. Various
attempts arrived to perform this measurement in one
dimension\cite{Soffitta95} or in two dimensions but at higher
energies\cite{austin94}. Eventually the Micropattern Gas Chamber
was developed with a full capability to convert photons and image
photoelectrons, making possible focal plane photoelectric
polarimetry.

\section{A NEW PATH TO X-RAY POLARIMETRY}
\subsection{The Micropattern Gas Chamber} The Micro Pattern Gas
Chamber\cite{costa01}$^,$\cite{Bellazzini06}$^,$\cite{Baldini06}
is a gas detector with a window, a drift region, a plane, finely
subdivided, electron multiplier (GEM) and a finely subdivided
array of metal pads collecting the charge. Pads are disposed with
hexagonal pattern on the top layer of a VLSI chip. Each pad is
equipped with a complete analog electronic chain so that the
charge collected is measured by a low noise electronics. The track
produced by the photoelectron that ionizes the gas are, therefore,
imaged by the device. From the analysis of the image the various
features of the track are reconstructed and the impact point is
determined with a resolution of the order of 150 $\mu$m FWHM. The
original direction of the photoelectron is reconstructed by
fitting the part of the track close to the impact point. The VLSI
chip is the core of the device. Three version have been realized
in rapid sequence. The last one is extremely
evolved\cite{Bellazzini06a}$^,$\cite{Bellazzini06b} and can be
foreseen, as it is or with minor adjustments, as the core of a
detector for a new mission of X-ray Polarimetry. This last version
of the device also manages the problems deriving from the very
large number of pixels (105600). The original version used the
signal from the GEM as a trigger and routed the holded analog
signals from each pad to the external A/DC. The advantage with
respect to a CCD is a synchronous reading of the device but the
time needed to convert the whole image (or a predefined part of it
in a segmented configuration) is of the same order. With
increasing number of pixels this could result in a large dead
time. In the last version this is overcome with a auto-trigger
capability and with the routing of only the data within a window
defined every time around the pixels that triggered. The device is
described in detail in another paper in these same
proceedings\cite{bellazzini06c}. Here we want to stress the
reasons why this device makes astronomical X-ray polarimetry more
attractive than what was possible with the conventional devices.

\subsection{Generalities on a measurement of polarization}
\label{sect:title}

A polarimeter  responds to incoming radiation by assigning
out-coming photons to an angular channel. In practice each photon
is coming out at a certain azimuth angle and an histogram of
events per angular bin is built. In dispersive polarimeters (such
as Bragg) one angle is measured at a time. In non dispersive
polarimeters (such as a MPGC or a Compton polarimeter) all the
output channels are collected simultaneously. When 100$\%$
polarized radiation impinges on an ideal polarimeter the histogram
of angles can be fitted with a cos$^{2}$ law. In a real
polarimeter the modulation is lower and a constant term must be
added. This defines the modulation factor $\mu$, namely the
(N$_{max}$ - N$_{min}$)/(N$_{max}$ + N$_{min}$) in the angle
histogram for a 100$\%$ polarized beam.

If photons are distributed around the expected value according to
Poisson statistics the relevant figure is the Minimum Detectable
Polarization (MDP) (at a confidence level of 99 \%) :
\begin{center}

              MDP = $\frac{4.29}{\mu S}$ ($\frac{S+B}{T}$)$^{0.5}$
\end{center}

where S is the source counting rate, B is the background, T the
observing time and $\mu$ is the modulation factor. S is the
product of the area A, the efficiency $\varepsilon$, the time and
the source flux.

In a focal plane imaging polarimeter the collecting area is
provided by the telescope and the background rate is negligible
with respect to the source one.  This is also true for a focal
plane Bragg crystal but with a much larger efficiency due to the
larger band. Moreover the rotation needed for the Bragg is a major
complexity and requires and independent monitoring for variable
sources.

Scattering polarimeters are background dominated and the response
is strongly affected by systematics that can only be partially
removed with the unavoidable rotation.

MPGC are good imagers and perform simultaneously timing and
spectra. They are suitable for energy resolved polarimetry. The
response with energy of the photoelectric polarimeter can be tuned
with a proper choice of the filling mixture. To do this we must
keep in mind that the MDP is proportional to the factor of merit
$\varepsilon^{0.5}\times\mu$. The control of systematics seems
very good\cite{Bellazzini06a} but in general the instrument will
be more effective when the modulation factor is high.

%%-----------------------------------------------------------

%%%%%%%%%%%%%%%%%%%%%%%%%%%%%%%%%%%%%%%%%%%%%%%%%%%%%%%%%%%%%
\section{A PATHFINDER TO POLARIMETRY?} \label{sect:sections}

%%-----------------------------------------------------------
\subsection{Polarimetry with small telescopes}
Since the sensitivity is a matter of number of photons if we want
to observe small polarized fractions from faint sources a large
collecting area is needed.  The optimal application of
photoelectric polarimetry is aboard a mission with a large
telescope, such as the proposed XEUS, that would access sources
down to a fraction of milliCrab. But the last available data were
collected 30 years ago. The improvement from OSO-8 to XEUS is
several orders of magnitude. There is room for a pathfinder with
intermediate sensitivity. In a complex mission the share of time
for polarimetry will be necessarily small, while a dedicated
mission could invest al the time to this discipline. A peculiarity
of polarimetry is that since we must operate in conditions that
the counts from the source exceed largely the counts from
background in the PSF, the data collected with different
telescopes can be added without significant loss with the respect
to the same data collected with a single telescope. This allows to
achieve a relatively large collecting surface, with no need of a
long focal length with the substantial advantage of the use of a
much cheaper launcher. Moreover to study a sample of faint sources
(mainly extragalactic) or to make detailed studies of brighter
sources, a  certain number of long pointing can be foreseen.

%
%%-----------
\subsection{A possible solution: JET-X telescopes}
Following this concept we made a very preliminary design of a
mission, named POLARIX, conceived as a pathfinder of future
polarimetry missions following the guidelines:
\begin{itemize}
\item Low Costs.
\item Established technologies to save development time.
\item Re-use of existing hardware.
\item Compatibility with a small launcher (but a minimum length is
needed).
\end{itemize}

A possible solution is the use of a cluster of telescopes like
those developed for JET-X experiment.\cite{citterio96}$^,$
\cite{wells97} A twin telescope was built and tested for the
Spectrum X-Gamma Mission that was never completed to the launch.
The spare unit has been very successfully used for the SWIFT
Mission\cite{burrows03}. Each telescope consists of 12 nested
double shells, with Wolter-1 geometry with an aperture of 300 mm
and a focal length of 3500 mm, manufactured with the
electroforming replica process. The measured resolution is better
than 15". The effective area of each telescope is around 160
cm$^2$ at 1.5 keV and has a response relatively flat in energy
allowing for good measurement up to 10 keV. The flight units of
JET-X could be a part of POLARIX. A relevant part of the
production costs of the telescope is the manufacture of
superpolished mandrels. Since they are still available we assumed
that 3 more telescopes could be built and integrated on POLARIX.
The cluster of telescopes would have a total area of 800 cm$^2$.
The total length is below 4 meters, compatible with low cost
launchers.

%%%%%%%%%%%%%%%%%%%%%%%%%%%%%%%%%%%%%%%%%%%%%%%%%%%%%%%%%%%%%
\section{Expected performances in a baseline configuration}

Let us assume to 5 JET-X like telescopes and in the focus of each
a MPGC with Ne (40$\%$) DME (60$\%$) filling.

   \begin{figure}
   \begin{center}
   \begin{tabular}{c}
   \includegraphics[height=10cm]{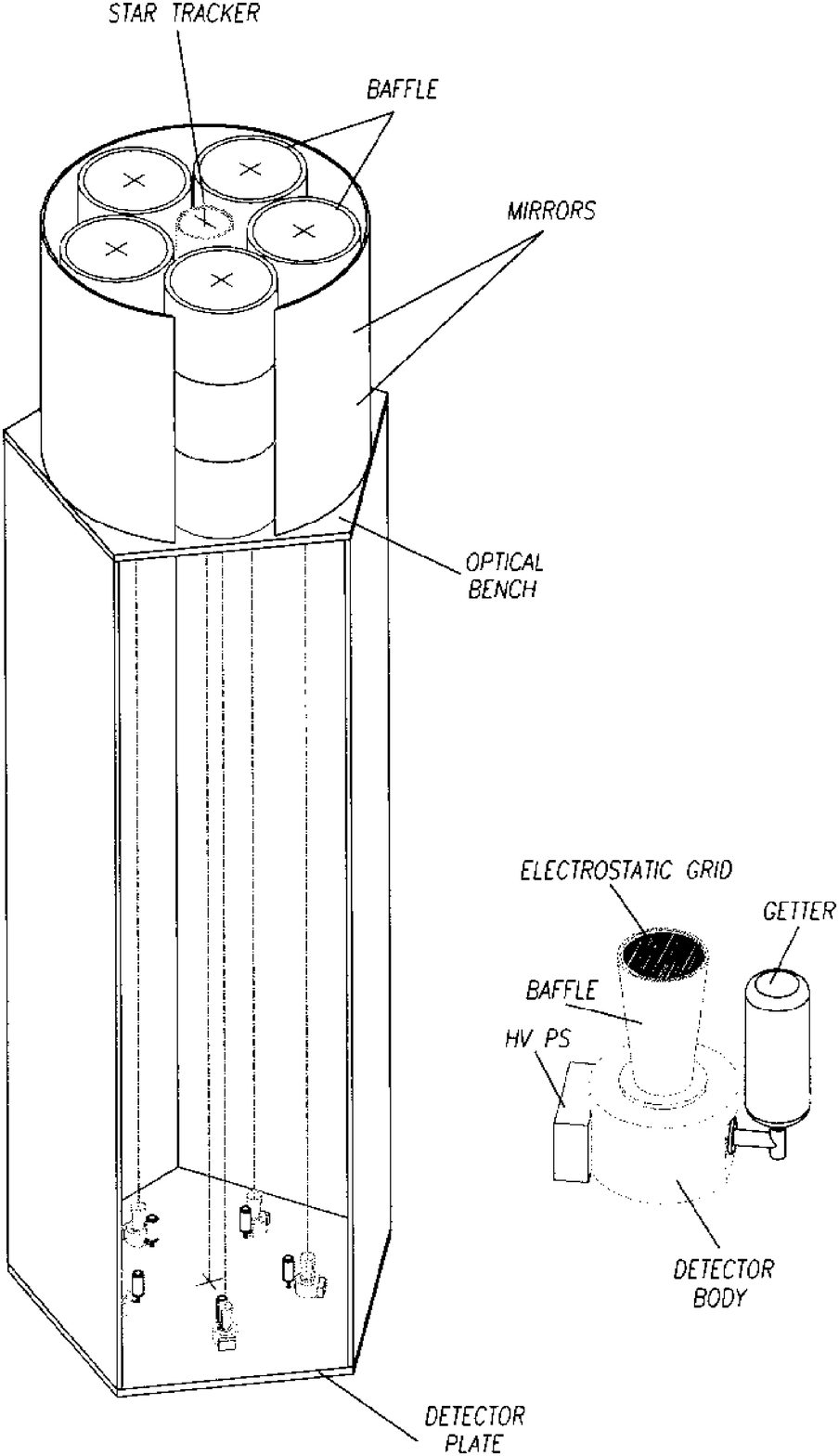}
   \end{tabular}
   \end{center}
   \caption[POLARIX]
%>>>> use \label inside caption to get Fig. number with \ref{}
   { \label{fig:POLARIX}
POLARIX.}
  \end{figure}

Beryllium is optimal for a sealed cell, without any gas flow
system, and can benefit of a long experience for space based
proportional counters. We assume to have a beryllium window 50
$\mu$m thick. This is the minimum value that does not requires a
back-skeleton that would reduce the effective area.

The MPGC beside being a polarimeter is an excellent imaging
device. The above mentioned capability to reconstruct the impact
point converts into a space resolution of around 150 $\mu$m FWHM.
This is intrinsically better than the telescope performance. But
we must consider another effect: the photons impinge inclined on
the detector and they are absorbed at different heights, according
to an exponential law. At the higher energies they, in practice,
are uniformly distributed, while at lower energies they are
absorbed in the majority closer to the window. If we mount the
detector in such a position that the focal plane is at half-way
from the window to the GEM (roughly a good choice but not
necessarily the best), the focal spot will be blurred because of
the projection on the detection plane of the interaction points.
We evaluate a worsening of the total response from 15 arcseconds
due to the telescope alone to $\sim$ 20 arcseconds from the two
combined effects.

This resolution is suitable for the angular resolved polarimetry
of a few but prototypic sources. In the case of Crab POLARIX is
capable to perform separate polarimetry of each of the two jets,
of inner torus plus the pulsar and of various parts of the outer
torus.

Time resolution can be very high in theory. Without any special
provision it can be fixed at 2 to 4 $\mu$sec. For a majority of
targets this will be redundant, due to the limited photon
statistics but for a few targets can be important.

Energy resolution will be that of a good proportional counter.
Depending on the mixture it can span from 10 to 14 $\%$ at 6 keV,
with a E$^{0.5}$ dependence. This is suitable for any reasonable
energy resolved measurement on the continua and, also, to separate
the unpolarized fluorescence from the partially polarized
continuum in reflection spectra.

As POLARIX would be a mission dedicated to the polarimetry in the
evaluation of the sensitivity we can also assume long pointing. We
stress that for an observation of 10$^{6}$ seconds the background
counting rate is still negligible even for the faintest sources
here considered. This is true for the statistics but, of course,
the background, that is mainly generated by gamma rays deriving
from an asymmetric source (cosmic, earth albedo, conversion of
Cosmic Rays in the spacecraft) could introduce some systematics to
be carefully kept under control.

\begin{figure}
   \begin{center}
   \begin{tabular}{c}
   \includegraphics[height=10cm, angle=90]{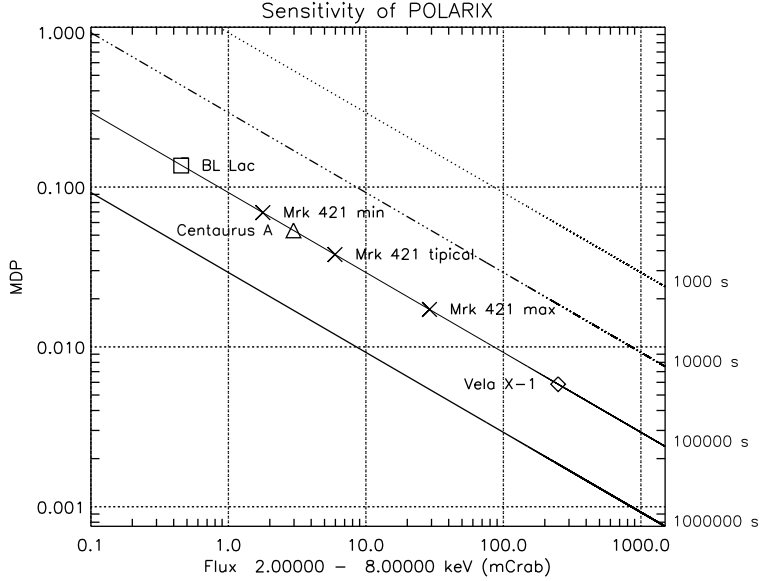}
   \end{tabular}
   \end{center}
   \caption[polarixmdp]
%>>>> use \label inside caption to get Fig. number with \ref{}
   { \label{fig:polarixmdp}
Minimum detectable polarization for POLARIX. A few representative
sources are shown on the sensitivity line with a pointing of
10$^{5}$sec.  }
  \end{figure}

In fig. \ref{fig:polarixmdp} we show the Minimum Detectable
Polarization for various sources of potential interest.

\begin{figure}
   \begin{center}
   \begin{tabular}{c}
   \includegraphics[height=10cm]{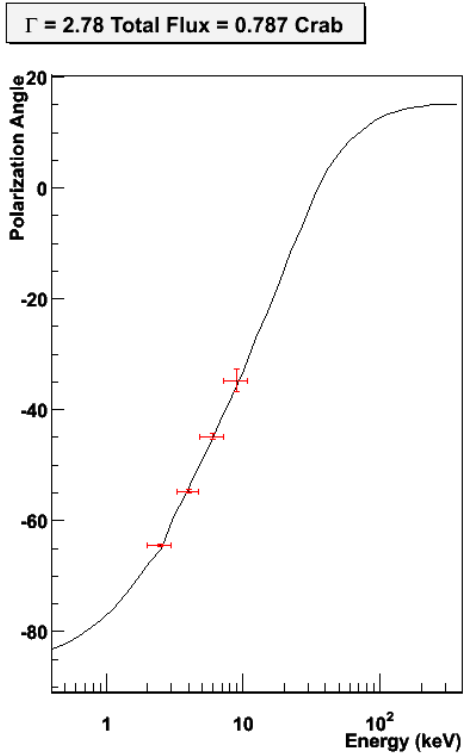}
   \end{tabular}
   \end{center}
   \caption[anglecy]
%>>>> use \label inside caption to get Fig. number with \ref{}
   { \label{fig:anglecy}
The rotation of the angle of polarization for Cyg X-1 as function
of the energy has been proposed for long time as a test of General
Relativity effects in the matter around a Black
Hole\cite{Connorsstark77}. In this figure we show the results of a
simulation of the capability of POLARIX to measure the
polarization angle as a function of energy, with a long
observation. A polarization of 5$\%$ has been assumed.The very
long exposure is needed only for the higher energy band}
  \end{figure}

\begin{itemize}
\item X-ray binaries can be studied in great details. POLARIX
can perform phase and energy resolved polarimetry of pulsators,
unveil the nature of the beam (fan or pencil), determine the
orientation of the rotation axis projected on the sky and directly
measure the angle between this and the magnetic field.
\item For Low Mass Binaries the polarization from scattering on
the disk can be determined.
\item The same for Black Hole Binaries for which the effects of
General Relativity should be detectable. As suggested by Stark and
Connors 30 years ago\cite{Connorsstark77} photons at different
energies are mainly generated at different radii in the disk. In
their path to the observer they will therefore be subject to a
different rotation of the polarization angle due to the different
gravity of the BH according to  General Relativity. This has been,
for long time, foreseen as the best evidence of the existence of a
Black Hole. In fig. \ref{fig:anglecy} we show the results of a
simulation of a measurement of 14 days  of Cyg X-1 with POLARIX.
We plot the expected dependence of the polarization angle on the
energy and we show the sensitivity of the measurement to detect
such a change in angle for a polarization of 5$\%$. For lower
polarization the sensitivity to the angle scales as the square
root of the polarized fraction. It is evident that even with a
polarization of the order of 2$\%$ the measurement is feasible.
\item The X-ray emission of most Blazars has been characterized by
various X-ray missions and is assumed to be well understood when
combined with observations at shorter and longer wavelengths. If
in the X-ray band the synchrotron is dominant a very high
polarization is expected. If the blazar is in such a state that
inverse Compton is the main source of the radiation the
polarization should be much lower or even null if the target
photons have a wide angular distribution (e.g. if they come from
the disk or from cosmic background). A few measurements from
POLARIX can test the unified model of blazars and enlighten the
physics of the jet.
\item In a few cases (at least in the subsystems of the Crab
Nebula but, likely, also in the brightest shell-like SNR) the
amount and orientation of regions of non thermal emission can
unveil the sites of acceleration of cosmic rays.
\end{itemize}

%%  Use following command to specify that graphics file is in
%%  a directory other than this LaTeX source file.
%%  Note use of / to separate subdirectories, for UNIX and Windows OS.
%%\graphicspath{{H:/HANSON/SPIESTY/}}
%-------------
%-------------
%%%%%%%%%%%%%%%%%%%%%%%%%%%%%%%%%%%%%%%%%%%%%%%%%%%%
%\appendix    %>>>> this command starts appendixes
%%%%%%%%%%%%%%%%%%%%%%%%%%%%%%%%%%%%%%%%%%%%%%%%%%%%
%\section{MISCELLANEOUS FORMATTING DETAILS} \label{sect:misc}

%It is often useful to refer back (or forward) to other sections in the article.  Such references are made by section number.  When a section reference starts a sentence, Section is spelled out; otherwise use its abbreviation, for example, ``In Sect.~2 we showed..." or ``Section~2.1 contained a description...".  References to figures, tables, and theorems are handled the same way.

%At the first occurrence of an acronym, spell it out, followed by the acronym in parentheses, e.g., noise power spectrum (NPS).

%%-----------------------------------------------

\section{Possible Improvements and trade-off.}

\subsection{Optics}

Relax the resolution of the telescopes to save weight. The JET-X
telescopes are based on the nickel replica technology and have
been designed to provide the excellent resolution of 15", needed
for the original application to imaging. They are relatively
heavy: each modulus weights 70 kg. For polarimetry sensitivity the
only requirement is that the background within the PSF is
(quadratically) negligible with respect to the source. This
situation will hold also if we relax the angular resolution. With
thinner nickel shells the resolution will degrade from 15" to 30"
and the total weight will decrease of more than 50$\%$. This could
be a good solution, at least for the telescopes which are still to
be manufactured. The only degradation of performances would be in
the domain of angular resolved polarimetry (where already a
degradation to 20" was expected due to inclined penetration of
photons), especially on Crab, but the trade-off could be
acceptable if it increases the feasibility of the mission.

A minor mismatching of MPGC with gold coated optics is the fact
that due to the M absorption edges of Au there is a drop in
reflectivity at energies where the polarimeter is particularly
sensitive. Pareschi et al\cite{pareschi04} have recently shown
that a thin Coating of Carbon, overimposed to the gold coating,
can fill this gap and increase the effective area in the 2 - 3 keV
band.\cite{pareschi04}$^,$\cite{pareschi04b} This possibility will
be seriously considered.

  \begin{figure}
   \begin{center}
   \begin{tabular}{c}
   \includegraphics[height=10cm, angle=90]{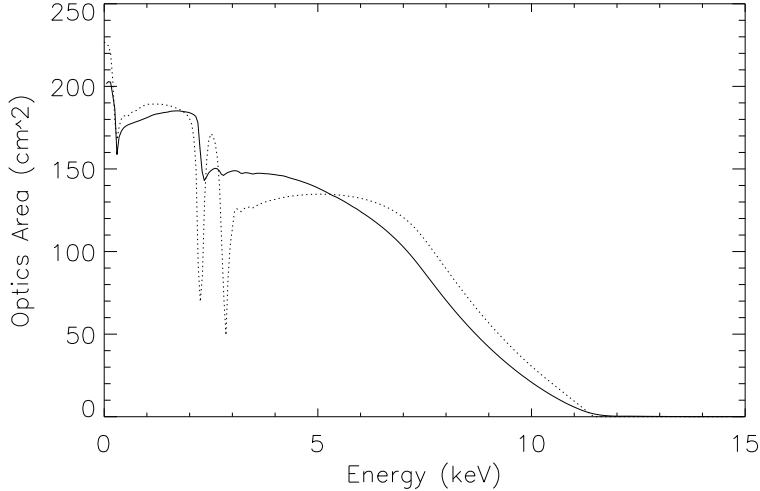}
   \end{tabular}
   \end{center}
   \caption[Carbon]
%>>>> use \label inside caption to get Fig. number with \ref{}
   { \label{fig:Carbon}
Effective Area of one JET-X modulus without the carbon coating and
with carbon coating.}
  \end{figure}

\subsection{Detectors}
In a MPGC the main trade-off is that on the gas mixture and on the
absorption gap thickness. Increasing the thickness or the
pressure, the efficiency increases and the modulation factor
decreases. This is further complicated by the fact that the
effective area of the telescope is another relevant parameter with
a fast dependence on the energy. Last but not least the fluxes
from the sources are usually decreasing with energy with a fast
law (power law or exponential). In practice even though the
polarimeter has a bandwidth much larger than a Bragg polarimeter,
the band is limited by these trade-offs. The choice of the optimal
band is not only a matter of technology but also a matter of
scientific preferences. Some targets, such as isolated pulsars or
Blazars in the synchrotron regime can be studied better at
energies as low as possible because of their spectrum. Sources for
which the polarization is expected to come from scattering, such
as low mass binaries, Seyfert Galaxies and QSOs, could be studied
much better at higher energies, where the scattering prevails on
photoabsorption. Also the phenomenology of X-ray pulsators is
expected to be more significant closer to resonance frequencies.
In some cases the dependence of polarization on energy is the most
valuable information, as in Black Hole binaries or blazars in the
transition from synchrotron to inverse Compton.

\begin{figure}
   \begin{center}
   \begin{tabular}{c}
   \includegraphics[height=13cm, angle=90]{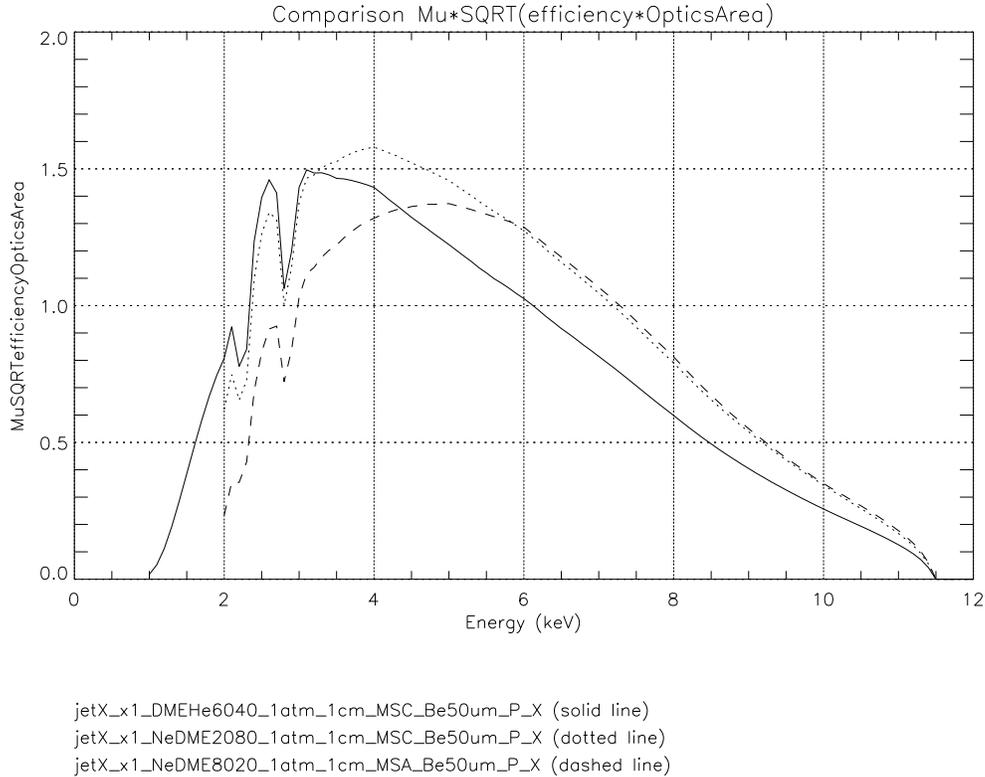}
   \end{tabular}
   \end{center}
   \caption[example]
%>>>> use \label inside caption to get Fig. number with \ref{}
   { \label{fig:example}
The factor of merit for three different mixtures tuned on
different bands.}
  \end{figure}

This opens the possibility that instead of having the same
detector in the focus of all telescopes we can use different
detectors, each tuned to a specific energy band. In fig.
\ref{fig:example} we show the $\mu \times \sqrt{\epsilon\times
Area}$ which is the factor of merit as a function of the energy
for three mixtures. The He/DME mixture optimizes the low energy
performance. This could be further stressed by using a filling gas
at a pressure below one atmosphere and a thin plastic window. The
Ne/DME (80$\%$-20$\%$) is more sensitive to higher energies. This
could be stressed with an overpressure or with a thicker
absorption gap, losing something at the low energy side, where the
modulation factor would decrease. A systematic study with a set of
benchmark sources is required to find the optimal configuration.
%-------------

\section{CONCLUSIONS}
POLARIX is a based on a new but already reliable technology and is
capable to perform a set of measurements of high astrophysic
interest and open the path to polarimetry with future large
aperture telescopes. An advanced study to evolve this concept to a
better defined design is about to start under ASI contract.

\section{Acknowledgments}
This research is sponsored by INFN,INAF and ASI
%%%%% References %%%%%

\bibliography{polreport}   %>>>> bibliography data in report.bib
\bibliographystyle{spiebib}   %>>>> makes bibtex use spiebib.bst

\end{document}